\documentclass[preprint,aps,nofootinbib]{revtex4}
\usepackage{dcolumn}
\usepackage{bm}
\usepackage{epsfig}
\usepackage{graphicx}
\usepackage{amsmath}
\usepackage{amssymb}
\usepackage{mathrsfs}
\usepackage{color}
\usepackage[usenames,dvipsnames]{xcolor}
\usepackage{hyperref}
\usepackage[caption=false]{subfig}
\usepackage[normalem]{ulem}

\def\be{\begin{equation}}
\def\ee{\end{equation}}
\newcommand{\beq}{\begin{equation}}
\newcommand{\eeq}{\end{equation}}
\def\bea{\begin{eqnarray}}
\def\eea{\end{eqnarray}}

\newcommand{\gsim}{ \mathop{}_{\textstyle \sim}^{\textstyle >} }
\newcommand{\lsim}{ \mathop{}_{\textstyle \sim}^{\textstyle <} }

\begin{document}

\begin{flushright}
\text{\normalsize MCTP-17-07}\\
\text{\normalsize WSU-HEP-1710}\\
\end{flushright}
\vskip 45 pt

\title{Stop Co-Annihilation in the Minimal Supersymmetric Standard Model Revisited}
\author{\bf \mbox{Aaron Pierce$^{\,a}$, Nausheen R.~Shah${\,^b}$ and Stefan Vogl${\,^c}$}}
\affiliation{
$^a$\mbox{\footnotesize{Michigan Center for Theoretical Physics, Department of Physics, University of Michigan, Ann Arbor, MI 48109}}\\
$^b$\mbox{\footnotesize{Department of Physics \& Astronomy, Wayne State University,  Detroit, MI 48201}}\\
$^c$\mbox{\footnotesize{Max Planck Institute for Nuclear Physics, 69117 Heidelberg, Germany}}\\
}

\begin{abstract}
\vskip 15 pt
We reexamine the stop co-annihilation scenario of the Minimal Supersymmetric Standard Model, wherein a bino-like lightest supersymmetric particle has a thermal relic density set by co-annihilations with a scalar partner of the top quark in the early universe.  We concentrate on the case where only the top partner sector is relevant for the cosmology, and other particles are heavy. We discuss the cosmology with focus on low energy parameters and an emphasis on the implications of the measured  Higgs boson mass and its properties. We find that the irreducible direct detection signal correlated with this cosmology is generically well below projected experimental sensitivity, and in most cases lies below the neutrino background. A larger, detectable, direct detection rate is possible, but is unrelated to the co-annihilation cosmology.
LHC searches for compressed spectra are crucial for probing this scenario.
\end{abstract}
\thispagestyle{empty}


\maketitle
\newpage

\section{Introduction}
The Minimal Supersymmetric Standard Model (MSSM) is a leading candidate for physics beyond the Standard Model~(SM). However, superpartners have remained stubbornly absent at the Large Hadron Collider (LHC).  Moreover, bounds from direct detection experiments, including LUX, PandaX, and Xenon1T \cite{Akerib:2016vxi,Tan:2016zwf,Aprile:2017iyp} present an increasingly strong challenge to the WIMP (weakly interacting massive particle) paradigm both in the MSSM and more broadly \cite{Kearney:2016rng}.  These experiments place the most pressure on models where dark matter--nucleon scattering is related to the cosmological history via a crossing symmetry.  Crossing symmetry is spoiled if the dark matter coinhabits 
the thermal bath with another exotic state at the time of freeze-out \cite{Griest:1990kh}. Then processes involving this co-annihilating partner can be important for the determination of  the relic density, but are unrelated (at tree level) to direct detection. These co-annihilating scenarios are therefore among the WIMP models least constrained by direct detection bounds. The MSSM realizes this scenario when the superpartner of the top quark, the stop, is light, and neutralino co-annihilations with this state determine the relic density.

In this work, we examine the stop co-annihilation scenario in terms of the low energy parameters most relevant for cosmology and direct detection. Many analyses of the stop co-annihilation parameters place an emphasis on simplified high-energy models, see e.g. Ref.~\cite{Ellis:2001nx,Ellis:2014ipa,Athron:2017yua,Athron:2017qdc}  However, because there are typically a small number of processes that dominate the cosmology involving
only a handful of  particles and couplings, it is illuminating to analyze these models in terms of the low energy parameters.  Consistent with this approach, in previous work \cite{Ibarra:2015nca} we considered the possibility where a single top partner (perhaps the superpartner of the right-handed top) was responsible for co-annihilation, see also Refs.~\cite{Delgado:2012eu,deSimone:2014pda,Garny:2012eb,Garny:2014waa,Kilic:2015vka, Berlin:2015njh} for related work. In the context of the MSSM,  however, a simplified model that includes a single co-annihilator, e.g. a $\tilde{t}_R$, may be too simple to capture the physics of both cosmology and direct detection. Mixing between the light stop and the heavy stop can impact both of these processes.  Indeed, the measurement of the Higgs boson mass suggests that there may be large mixing in the stop sector.  In this case, it makes sense to include the full stop sector in the simplified model.

This article is organized as follows. In Sec.~\ref{s.orientation} we discuss the basics of the parameter space and discuss constraints unrelated to the dark matter story. Cosmology is analyzed 
in  Sec.~\ref{s.relic} and we turn to implications for direct detection in Sec.~\ref{s.DD}.  Finally, Sec.~\ref{s.conc} is reserved for our conclusions. 
  
\section{Orientation}\label{s.orientation}

We consider the case of  pure bino neutralino dark matter $\chi$, with co-annihilations resulting from the presence of one or more colored states $Y$ with masses not dissimilar to that of the neutralino. If equilibrium between the neutralino and the colored state is maintained (as is typically the case due to scatterings off the SM bath), then processes of the form $(\chi Y \rightarrow SM)$ or $(YY \rightarrow SM)$ are relevant for setting the dark matter abundance.

A small admixture of Higgsino will not affect the cosmological history in detail.  However, doping the bino with even a small Higgsino fraction can impact direct detection since the Higgs boson has a tree-level bino-Higgsino coupling. The Higgsino fraction of the neutralino is controlled by mixing suppressed by $M_{Z}/\mu$, with $\mu$ the Higgsino mass parameter. The tree-level direct detection cross section is well approximated by \cite{Ibarra:2015nca}: 
\be\label{e:HDD}
\sigma_{SI}^{tree-mixing} \approx 3 \times 10^{-48}\textrm{cm}^2 \left( \frac{2 \textrm{ TeV}}{\mu} \right)^4 \left( \frac{m_\chi}{600 \textrm{ GeV}} \right)^2 \left(1+\frac{\mu \, s_{2 \beta }}{m_{\chi }}\right)^2 \left(1-\frac{m_{\chi }^2}{\mu ^2}\right)^{-2}.
\ee
Here, $s_{2 \beta} \equiv {\sin 2 \beta}$ where $\tan \beta$  is the ratio of vacuum expectation values of the two Higgs doublets.  In this work, we will be interested in the case where this contribution to direct detection is subdominant.  That is, we  consider the case where $\mu$ is (quite) large, and we ask the question:  what is the direct detection cross section induced only by the presence of a stop co-annihilation cosmology?  

Because we will be interested in discussing the effects of reproducing the required Higgs boson mass, we will include the full stop sector, i.e both $\tilde{t}_R$ and $\tilde{t}_L$.  This means that the left-handed sbottom $\tilde{b}_L$ is necessarily in the spectrum as well.  For simplicity, we assume the partner of the right-handed bottom is decoupled.  The precise value of its mass has little effect on either the physical Higgs boson mass or any fine-tuning arguments.\footnote{While we do not focus on this case, it is possible that the right-handed sbottom could be responsible for co-annihilation with the neutralino.  We will comment briefly on this case in Sec.~\ref{s.DD}.} The stop mass matrix is given by:
\begin{equation}
M^2_{\tilde{t}}=\left(
\begin{array}{cc}
m_{Q_3}^2 + m_t^2 + D_L   & \quad m_t X_t \\ 
& \\
m_t X_t
&  \quad m_{u_3}^2 + m_t^2 + D_R\\
\end{array}
\right),
\end{equation}
where $X_t = (A_t -\mu/\tan\beta)$,  $m_{Q_3},m_{u_3}$ are the left and right-handed soft masses, and the $D$-terms are: $D_L=  M_Z^2 \cos 2 \beta (\frac{1}{2}-\frac{2}{3} s_W^2)$ and $D_R=\frac{2}{3} M_Z^2 \cos 2\beta s_W^2$.  Here, $s_{W} = \sin \theta_W$ is the weak mixing angle.  We choose sign conventions for the sign of the stop mixing angle consistent with {\tt{SuSpect\_2.41}}~\cite{Djouadi:2002ze}.  
We have:
\begin{equation}
\sin 2\theta_{\tilde{t}} \equiv s_{2t}= \frac{-2 m_t X_t}{m_{\tilde{t}_2}^2-m_{\tilde{t}_1}^2   }\;,
\end{equation}
where we denote the stop mass eigenstates as   
$\tilde{t}_1 = c_t \, \tilde{t}_L +s_t \, \tilde{t}_R$ and $\tilde{t}_{2} =-s_t \, \tilde{t}_{L} +c_t \, \tilde{t}_R$  and use the shorthand $s_{t}= \sin \theta_{\tilde{t}}$, etc.  The stop mixing angle is a function of $X_t$ rather than $A_t$, $\mu$ or $\tan \beta$ directly. In the region of interest, $X_t \sim A_t$, and the dependence on $\tan \beta$ is minimal. For concreteness we fix 
$\tan\beta=10$ but numerically verify that results are approximately independent of 
$\tan\beta$, as long as the values are not too extreme. 

 The heaviness of the right-handed sbottom ensures that the $\tilde{b}_1$ state has mass given approximately by $m_{Q_3}$. But for simplicity,  we also set $X_b=0$ so there is no mixing in the sbottom sector, regardless of $\tan \beta$.  This ensures that the  $\tilde{b}_1$ is  completely left-handed, and its mass may be expressed in terms of the stop masses and mixing angles:
\begin{equation}
m_{\tilde{b}_1}\equiv m_{\tilde{b}_L} = c_t^2 m_{\tilde{t}_1}^2+s_t^2 m_{\tilde{t}_2}^2-M_W^2 c_{2\beta} +m_b^2-m_t^2\;.
\end{equation}

For later reference, we record the couplings of the stops and the left-handed sbottom to the Higgs and the bino: 
\begin{eqnarray}
\label{ht1t1}
g_{h\tilde{t}_1\tilde{t}_1}&=& -\frac{2}{v}\left[ M_Z^2 c_{2\beta}\left( \frac{1}{2} c_t^2 -\frac{2}{3} s_w^2 c_{2 t}\right) +m_t \left(m_t+\frac{1}{2} s_{2 t} X_t\right) \right] \; ,\\
\label{ht2t2}
g_{h\tilde{t}_2\tilde{t}_2}&=& -\frac{2}{v}\left[ M_Z^2 c_{2\beta}\left( \frac{1}{2} s_t^2 +\frac{2}{3} s_w^2 c_{2 t}\right) +m_t \left(m_t-\frac{1}{2} s_{2 t} X_t\right) \right] \; , \\
\label{ht1t2}
g_{h\tilde{t}_1\tilde{t}_2}&=& \frac{2}{v}\left[ M_Z^2 c_{2\beta}s_{2 t} \left( \frac{1}{4} -\frac{2}{3} s_w^2 \right) -\frac{1}{2} m_t c_{2 t} X_t \right] \; ,\\
\label{t1xx}
g_{\tilde{B}\tilde{t}_1{t}}&=&\frac{\sqrt{2}}{3} g_Y \left(-\frac{1}{2}c_t P_L + 2s_t P_R\right)\; ,\\
\label{t2xx}
g_{\tilde{B}\tilde{t}_2{t}}&=& \frac{\sqrt{2}}{3} g_Y \left(\frac{1}{2}s_t P_L + 2c_t P_R\right)\; ,\\
\label{bLx}
g_{\tilde{B}\tilde{b}_L{b}}&=& - \frac{\sqrt{2}}{6}g_Y \; ,
\end{eqnarray}
where the Higgs vacuum expectation value $v=246$ GeV, $g_Y$ is the hypercharge gauge coupling and $P_L$, $P_R$ are the projection operators.

We implemented this simplified model containing the third generation squark doublet, right-handed stop  and a bino neutralino  in {\tt MicroOmegas}~\cite{Belanger:2013oya}. 
As a cross-check we validated our numerics with the full MSSM implementation.    

\subsection{ Indirect Constraints}\label{s.Higgs}

Under the assumption that the MSSM is the underlying theory, important information about the superpartner spectrum can be gleaned from the Higgs boson mass. Indeed, the MSSM cannot reproduce a Higgs boson mass of $125$ GeV without substantial radiative corrections from the stop sector. Because the Higgs mass constraint impacts the allowed properties of the stops, it is of interest to study whether it has phenomenological implications for the stop co-annihilation region.  Nevertheless, one should bear in mind the possibility that the observed Higgs mass is generated by physics beyond the MSSM.  

We therefore prefer to begin by considering additional indirect consequences of the stop sector, potentially relevant even if there are additional contributions to the mass of the Higgs boson.  In particular,  before discussing the impact of the Higgs mass, we discuss three sets of constraints: electroweak precision observables (EWPO),  Higgs boson production and decays, and the stability of the physical vacuum due to the presence of charge and color breaking vacuua.  It is a model-dependent question as to whether the above 
would be affected by deviations from the MSSM.  At minimum, however, these constraints  give an indication of cancellations that would need to occur between the stop sector and any additional contributions.  Once collider bounds are imposed, we find constraints from vacuum stability are generally the strongest of the three.

\begin{figure}[tb]
\begin{tabular}{cc}
 \includegraphics[width=0.5\textwidth]{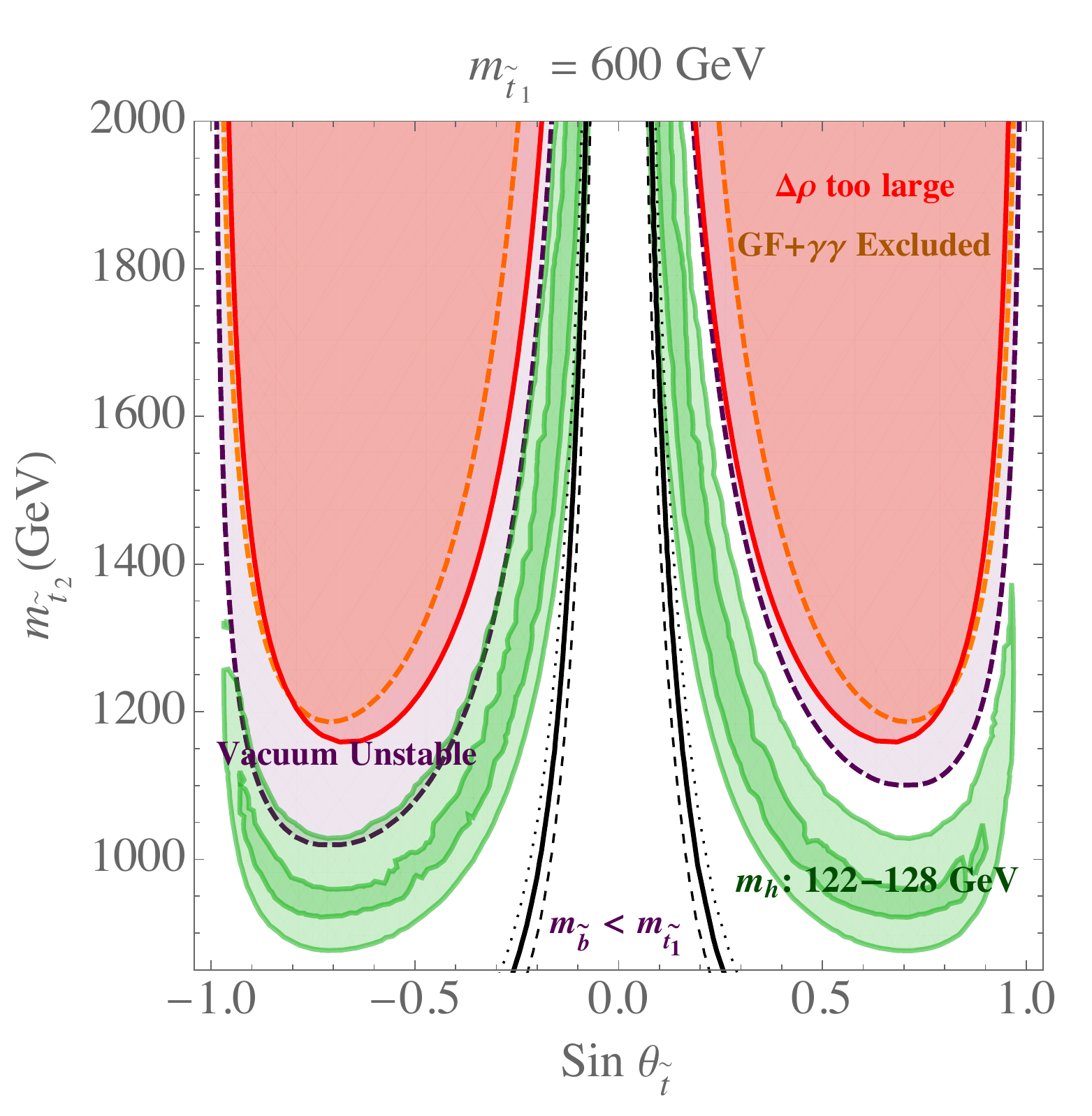}~~ & ~~
 \includegraphics[width=0.5\textwidth]{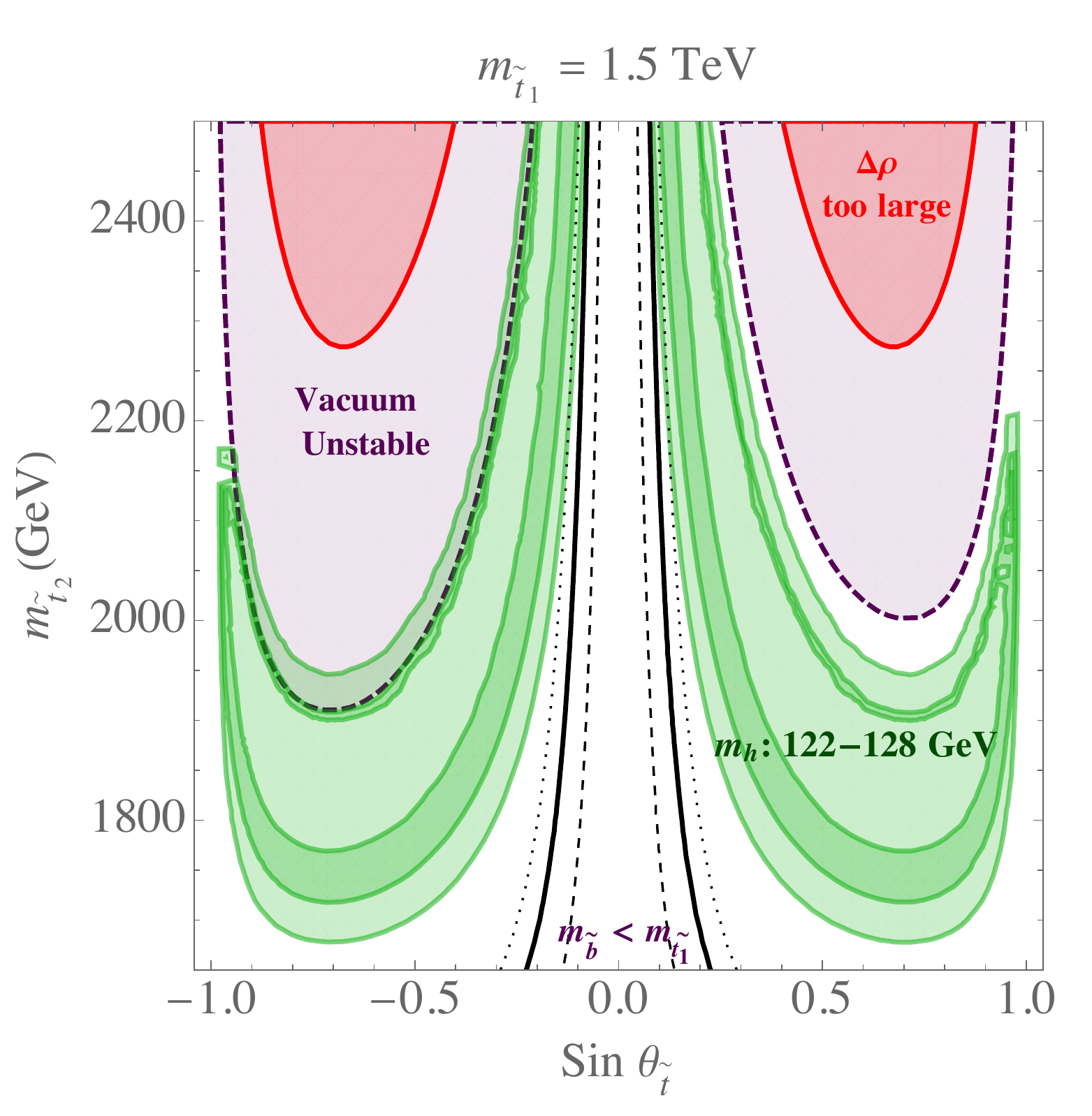} \\
 (a) ~~ & ~~ (b)
\end{tabular}
\caption{ \label{fig:Higgs} {Impact of $\delta \rho$~(excluded red region), vacuum stability~(excluded purple region) and Higgs phenomenology~(excluded orange region) on the $\tilde{t}$ sector. Also shown in green is the region that yields a observationally consistent Higgs boson mass $122 < m_{h} < 128$ GeV; there are two overlapping regions  corresponding to different sign choices for the gluino mass.  {\it Left: $m_{\tilde{t}_1}= 600 $} GeV, {\it Right: $m_{\tilde{t}_1}= 1.5 $  }TeV.    Also shown are contours where the sbottom is degenerate with the stop (solid) or within $\pm 5$ GeV [dashed (-) / dotted (+)].}}
\end{figure}

Regarding EWPO, we require the stop sector contribution to $\delta \rho$ not exceed the $2\sigma$ bound from the measured $\delta \rho = 3.7  \pm 2.3 \times 10^{-4}$  \cite{Olive:2016xmw}. Assuming negligible mixing in the sbottom sector, the third generation squarks yield \cite{Bilal:1990gh}:
\begin{equation}
\delta \rho
= \frac{3 G_{F}}{8 \sqrt{2} \pi^2} \left( -s_t^2 c_t^2 F_0[m_{\tilde{t}_1}^2,m_{\tilde{t}_2}^2] + c_t^2 F_0[m_{\tilde{t}_1}^2, m_{\tilde{b}_L}^2] + s_t^2 F_0[m_{\tilde{t}_2}^2, m_{\tilde{b}_L}^2]\right),
\end{equation}
with $G_{F}$ the Fermi constant 
while the loop function $F_0$ is given by 
\begin{equation}
F_0[x,y] \equiv x+ y - \frac{2 xy }{x-y} \log\left[\frac{x}{y}\right].
\end{equation}
In the limit of negligible mixing in the stop sector, the bound on $\delta \rho$ implies 
 $m_{\tilde{t}_L} > 290 $ GeV. In the case with hierarchical stops, where the lightest stop is still mostly left-handed,  it can be shown that $X_t\sim m_{\tilde{t}_2}$ minimizes the above contribution to $\delta\rho$~\cite{Batell:2013psa}. This is due to an approximate custodial symmetry between the sbottom and the lightest stop -- these two states have an approximate mass degeneracy in this regime: $m_{\tilde{t}_1}^2 \equiv m_{Q_3}^2 +m_t^2\left( 1-X_t^2/m_{u_3}^2\right) \sim m_{Q_3}^2\equiv m_{\tilde{b}_1}^2$.  Deviating too far from this limit (i.e. very large $X_{t}$), causes the constraints to become significant.  The excluded region is shown  
in Fig.~\ref{fig:Higgs} (red region) in the $\sin\theta_{\tilde{t}}$ vs. $m_{\tilde{t}_2}$ plane for two different values of  $m_{\tilde{t}_1} = 600$ GeV (a) and 1.5 TeV (b).

Higgs boson properties can also be influenced by the presence of a light stop.  
The primary impact
is a potential contribution to the gluon fusion production cross section.  The contribution to $H\rightarrow \gamma \gamma$, while less pronounced, should also be considered.
Using the Higgs low energy theorem, the contribution of $\tilde{t}_{1,2}$ to the production amplitude of the Higgs boson in gluon fusion can be approximated as~\cite{Kniehl:1995tn, Carena:2012xa, Carena:2013iba} 
\beq\label{e:gf}
\mathcal{A}_{hgg} \simeq \mathcal{A}_{hgg}^{\rm SM} +  \frac{m_t^2}{ m_{\tilde{t_1}}^ 2m_{\tilde{t_2}}^2}(m_{\tilde{t}_1}^2+m_{\tilde{t}_2}^2-X_t^2)\,,
\eeq 
where $A^{\rm{SM}}_{hgg} = 4$ is the SM amplitude.
Analogously we can write the amplitude for $H\rightarrow \gamma \gamma$ as 
\beq\label{e:gg}
\mathcal{A}_{h\gamma\gamma} \simeq \mathcal{A}_{h\gamma\gamma}^{\rm SM} +\frac{8}{9}  \frac{m_t^2}{ m_{\tilde{t_1}}^ 2m_{\tilde{t_2}}^2}(m_{\tilde{t}_1}^2+m_{\tilde{t}_2}^2-X_t^2)\,,
\eeq 
where $\mathcal{A}_{h\gamma\gamma}^{\rm SM} = -13$~~\cite{Kniehl:1995tn, Carena:2012xa, Carena:2013iba,Ellis:1975ap,Shifman:1979eb,Gunion:1989we}.  Again, if $X_{t} \gg m_{\tilde{t}_2}$, both gluon fusion and the diphoton rate will be impacted, but because $| \mathcal{A}_{hgg}^{\rm SM} | \ll |\mathcal{A}_{h\gamma\gamma}^{\rm SM}|$,
the stop contribution to gluon fusion provides a larger deviation from SM expectations.
The combined analysis of the 8 TeV LHC run by ATLAS and CMS constrains 
 the signal strength of the production in gluon fusion, normalized to the SM prediction, to be $\mu_{ggF} = 1.03^{+0.17}_{-0.15}$  while $\mu_{\gamma \gamma} = 1.16^{+0.20}_{-0.18}$~\cite{ATLAS-CONF-2015-044}.
 Allowing for near future improvements in these measurements, we allow a deviation of up to 20\% in these observables.  
  With the increasingly powerful direct searches for the stop (which impose $\tilde{t}_{1} \gsim 500$ GeV, more below), these constraints are only relevant for particularly large values of the $X_{t}$ parameter.   The region excluded by Higgs signal strength measurements is shown in orange in Fig.~\ref{fig:Higgs}.

Charge and color breaking vacua may appear at large values of the trilinear coupling $A_{t}$ and the physical vacuum can become metastable, possibly with a lifetime shorter than that of the observed universe.   To ensure a metastable physical vacuum with a sufficiently long lifetime, following Ref.~\cite{Blinov:2013fta}\footnote{We have verified with the authors of this reference that this approximation remains roughly applicable for the larger values of $\mu$ which we are interested in here.}  we impose the rule of thumb condition \begin{equation} \label{eq:At}
A_{t}^{2} \leq \left( \alpha + \beta \frac{|1-r|}{1+r}\right)(m_{\tilde{t}_1}^2 +m_{\tilde{t}_2}^2) - \gamma M_Z^2/2\; ,
\end{equation}
with $\beta = -0.5$, $\alpha= 3.4$, $\gamma =60$ and $r \equiv m_{u_3}^2/m_{Q_3}^2$.  
We see that
stability of the physical vacuum roughly requires $|A_{t}| \lesssim m_{\tilde{t}_2}$.  Because $X_t \equiv \left(A_t -\mu/\tan\beta\right)$,  the precise impact of Eq.~(\ref{eq:At}) on the stop sector will depend on the choice of $\mu$ and $\tan\beta$. However, even for fairly large values of $\mu$, as long as $\tan\beta$ is larger than a few, $X_t\sim A_t$ and the impact of the exact choice of $\mu$ and $\tan\beta$ is minimal. For concreteness, when imposing the vacuum stability constraint, Eq.~(\ref{eq:At}), on the stop parameters, we fix $\mu = 3~ m_{\tilde{t}_1} \sim 3~m_\chi$, a value large enough that  the tree-level direct detection cross-section will be small, see~Eq.~(\ref{e:HDD}).\footnote{In discussing direct detection in Sec.~\ref{s.DD}, however, we explicitly set bino-Higgsino mixing to zero (as would be appropriate for very large $\mu$),  focusing instead on the contributions arising from the stops/sbottom.}
In Fig.~\ref{fig:Higgs}, the purple region denotes where the vacuum becomes unstable according to this prescription. The slight asymmetry between positive and negative values of $\sin \theta_{\tilde{t}}$ is because of
the difference between $X_t$ and $A_t$ due to our choice of $\mu/\tan\beta$.  
 
 In the absence of Higgs mass information, the above constraints are responsible for limiting the maximal value of
 $X_t$ allowed. Because $X_{t}$ enters the Higgs boson couplings to the stops, this affects both cosmology and direct detection.  As we will now see, the Higgs mass constraint gives comparable, and often stronger bounds on $X_{t}$. 

\subsection{Higgs Mass}\label{s.Mass}
At tree level, the MSSM predicts a Higgs boson with mass 
below $M_{Z}$. However, loop effects give important contributions \cite{Ellis:1990nz,Haber:1990aw,Okada:1990vk}. In our analysis, we employ the Higgs mass calculation as implemented in {\tt FeynHiggs-2.13.0}~\cite{Bahl:2016brp, Hahn:2013ria,Frank:2006yh,Degrassi:2002fi, Heinemeyer:1998np,Heinemeyer:1998yj}. In the stop sector, we scan over $m_{Q_3}$, $m_{u_3}$ and $A_t$.  For concreteness we fix $M_3 = \pm 1.5$ TeV, $X_b=0$, and all other soft masses and trilinear couplings to 5 TeV. To capture potential uncertainties due to even higher order corrections and – more importantly – uncertainties in inputs (such as the top quark mass or the Gluino mass, which can change the Higgs mass up to 4 GeV when the Gluino mass is varied from 1 to 3 TeV \cite{Degrassi:2001yf,Bagnaschi:2014rsa}) we  allow a Higgs boson mass in a window from 122 to 128 GeV. The sign of $(A_t M_3)$ has an asymmetric effect on the Higgs mass primarily due to the running of the top Yukawa (see Ref.~\cite{Draper:2016pys} and references therein). However, this sign does not otherwise affect the dark matter phenomenology.  
Hence, we check both signs of the gluino mass, to see whether either realizes a valid Higgs mass. 

Consistency with the observed Higgs boson mass in the presence of a relatively light $\tilde{t}_1 \lsim $~TeV  requires a relatively heavy  $\tilde{t}_2$.  Unless one wishes to allow for an extreme hierarchy between the stop masses, an appreciable mixing, $|X_t| \sim m_{\tilde{t}_2}$ is also preferred~\cite{Carena:2011aa, Carena:2012gp,Carena:2013iba, Batell:2013psa}. 
In Fig.~\ref{fig:Higgs}, the green swath denotes the region with an observationally consistent Higgs boson mass. The two bands that comprise the swath correspond to the two signs of $M_3$. 

Also in Fig.~\ref{fig:Higgs}, we indicate where the sbottom mass is degenerate with $m_{\tilde{t}_1}$ by the solid black line.  The two dashed (dotted) contours show where the sbottom is 5 GeV lighter (heavier) than  $m_{\tilde{t}_1}$. The central region enclosed by the thick black lines is then where the sbottom is lighter than the lightest stop, and hence would be the next to lightest supersymmetric particle (NLSP).  Close to these lines, we expect the sbottom and stop to act as co-NLSPs.  This information will be relevant when we discuss the cosmology associated with this scenario.

 To summarize, the region compatible with the Higgs boson mass is mostly unaffected by the indirect constraints of the previous section.
 However,
 the largest values of $X_{t}$ consistent with the  Higgs mass can be in tension with the stability of the vacuum. 

\subsection{LHC}\label{s.LHC}

Owing to its small production cross section, pure bino dark matter is difficult to probe at the LHC.  The colored, nearly degenerate co-annihilation partner, on the other hand, possesses a relatively large production cross section.  While small mass splittings between the states degrade the classic missing energy signature, a variety of searches targeting this region
now exist.  We review a few of the most relevant searches here. Other particles in the SUSY sector are also potentially accessible, i.e. $\tilde{t}_2$ and $\tilde{b}_1$. Since they will typically have a larger splitting with the lightest supersymmetric particle~(LSP), the sensitivity to these states can be enhanced. Hence, in some cases it is possible that their limits can be important, despite the smaller production cross section. 

 In the co-annihilation scenario the mass splitting between the $\tilde{t}_1$ and the dark matter is always less than $m_{t}$, so the two-body decay of the stop $(\tilde{t_1} \rightarrow \chi^0 t)$ is kinematically forbidden. 
 Indeed, for all but the heaviest stops, the splitting is much less than $M_W$, which also forbids the three-body decay $(\tilde{t_1} \rightarrow \chi^0 W^+ b)$,  leaving  only the four-body decay $(\tilde{t} \rightarrow b \chi^{0} f \bar{f}')$, or the  flavor violating two-body decay $(\tilde{t} \rightarrow c \chi^{0})$.
 The relative importance of these last two decays depends on the flavor properties of the SUSY breaking sector~\cite{Blanke:2013uia,Agrawal:2013kha}.  Recent searches by the CMS collaboration have focused on this compressed region, and have ruled out stop masses up to roughly 500 GeV~\cite{CMS-PAS-SUS-16-049}.  The exact mass excluded depends on the precise value of the mass splitting and the relative branching ratio of these two channels. Monojet searches, such as those undertaken at ATLAS~\cite{Aaboud:2016tnv} can also be relevant, but typically are not as constraining.

For $m_{\tilde{b}} > m_{\tilde{t}_1}$ the sbottom decay pattern depends strongly on the mass splitting.  If kinematically accessible and if there is any left-handed component in the lightest stop,  ($\tilde{b} \rightarrow \tilde{t}  W$) dominates, which is challenging to observe.  However, a dedicated search may be possible.     If the splitting is less than $M_{W}$, $(\tilde{b} \rightarrow b \chi)$ robustly excludes  $ m_{\tilde{b}} < 625$ GeV ~\cite{CMS-PAS-SUS-16-032}.
A phenomenological projection~\cite{An:2016nlb} suggests that a combined search for all the different decay modes of the sbottom could ultimately probe $m_{\tilde{b}} \lesssim 900$ GeV with $300 \; \mbox{fb}^{-1}$ of data.

Finally, the decay width of the lightest stop is suppressed if the mass splitting is small, and for $\Delta m \lesssim 10$ GeV the $\tilde{t}_1$  has a proper decay length exceeding a meter \cite{Grober:2014aha}. Current limits on such long-lived particles exclude stable stops with $m_{\tilde{t}_1} < 1020$ GeV \cite{Khachatryan:2016sfv}.  As we shall see, a co-annihilation cosmology does not motivate such a small mass splitting for stops this light (see next section).  Even with 300 fb$^{-1}$ it will be challenging for  these searches to probe the relevant region, where the stop mass exceeds 1.5 TeV.

At present, stops with modest mass splittings from the neutralino and mass roughly greater than 500 GeV are allowed, so in what follows we will focus on this region.   The LHC will continue to push these bounds upwards as additional data are taken.

\section{Relic Density}\label{s.relic}

In the model studied here, co-annihilation of $\chi$ with $\tilde{t}_1$ and $\tilde{b}_1$~\cite{Ellis:2001nx}, as well as the annihilation of  pairs of $\tilde{t}_1$  and/or $\tilde{b}_1$  are relevant for the precise determination of $\Omega h^2 = 0.11805 \pm 0.0031$ \cite{PlanckCollaboration2013}. For some spectra, even processes involving $\tilde{t}_{2}$ can be relevant.
Contributions from processes with particles other than $\chi$ in the initial state must be appropriately weighted by their thermal abundance.   The impact of these additional processes is accounted for via an effective annihilation cross section \cite{Griest:1990kh} 
\begin{eqnarray}
\sigma_{eff} {\rm v} = \sum_{i,j} \frac{n_i^{eq} n_j^{eq}}{\left(\sum_k n_k^{eq}\right)^2}\sigma_{ij} {\rm v}\;,
\end{eqnarray}
where $n_i^{eq} = g_i [m_i T/ (2 \pi)]^{3/2} e^{-m_i/T}$; $m_i$ is the mass of the particle $i$, and $g_i$ counts the number of internal degrees of freedom. 

Since the abundance of a heavier state $i$ is suppressed relative to that of the LSP by factors of $e^{-\Delta m /T_F}$,  where $\Delta m = \left(m_{i} -m_{\chi}\right)$, the relic density is extremely sensitive to the mass splitting between $\chi$ and the co-annihilator. For $m_{i} \gtrsim 1.2\, m_{\chi}$ co-annihilations can safely be neglected. 
Contributions to the effective cross section from the annihilations of a pair of NLSPs are doubly exponentially suppressed compared to $\chi\chi$ annihilations, and the annihilations of  $\chi$ with a single co-annihilator are singly exponentially suppressed. Note,  strongly interacting particles $\tilde{f}$ always possess the annihilation channel $(\tilde{f} \tilde{f^{\ast}} \rightarrow gg)$ with cross section
\be
 \sigma {\rm v} \left(\tilde{f} \tilde{f}^*\rightarrow gg\right)= \frac{7 g_s^4}{216 \, \pi \,  m_{\tilde{f}}^2} \, .
 \ee 
Thus, co-annihilations impose a lower limit on $\Delta m$ as a function of $m_{\chi}$, since co-annihilations will be too effective  if the mass splitting is too small.\footnote{ This conclusion may be avoided if chemical equilibrium between the dark matter and its co-annihilation partner does not hold~\cite{Garny:2017rxs}, but for the current scenario processes such as $(\chi t \rightarrow \tilde{t} g)$ are expected to be sufficiently rapid to maintain chemical equilibrium.}

A complication arises from the Sommerfeld effect, which is known to have an impact on the annihilation rate of charged non-relativistic particles~\cite{Orig:Somm,Hisano:2003ec,Hisano:2004ds,ArkaniHamed:2008qn}.   While the Sommerfeld effect does not directly affect the annihilation of binos, 
it will have an impact on the annihilation rates of  $\tilde{t} \tilde{t}^{\ast}$ and  $\tilde{t} \tilde{t}$ (and potentially their analogs including the sbottom). 
These corrections can modify the relic density significantly~\cite{deSimone:2014pda,Baer:1998pg,Freitas:2007sa,Hryczuk:2011tq,Ibarra:2015nca}. Qualitatively, the Sommerfeld effect can be understood as an enhancement (or suppression) of the leading order cross section due to the presence of an attractive (repulsive) potential between the initial state 
particles generated by the exchange of light force carriers. In the context of color charged particles such a potential is generated by gluon exchange  and can be approximated by the leading term of the static QCD potential \cite{Fischler:1977yf},\cite{deSimone:2014pda}:
\begin{align}
V(r)\approx C \frac{\alpha_S}{r} = \frac{\alpha_S}{2 r} \left(C_Q-C_R-C_{R'}\right)\;,
\end{align}
where $C_R$ and $C_R'$ are the quadratic Casimirs of the color representation  of the incoming particles while $C_Q$ is the quadratic Casimir of the final state.
In this case the long and the short range contributions to the annihilation process factorize. The Sommerfeld corrected  s-wave cross section is given by $\sigma^S \equiv S_0 \;\sigma_{0}$ where  $\sigma_{0}$ is the short distance contribution, and $S_0$ is an enhancement (suppression) factor given by
\begin{align}
S_0=\frac{\pi \alpha/\beta}{ e^{\pi \alpha/\beta}-1}\;.
\end{align}  
Here $\alpha \equiv 1/2 \,\alpha_S \,(C_Q-C_R-C_{R'})$ is the strength of the potential, and $\beta=v/2$ where $v$ corresponds to the velocity of the particles in the initial state. The appropriate Sommerfeld factors for higher partial waves $S_l$ can be constructed using a recursion relation starting from $S_0$~\cite{Cassel:2009wt}. 
 For a detailed description of our treatment of the Sommerfeld effect see Ref.~\cite{Ibarra:2015nca}. 
We have implemented the Sommerfeld effect in {\tt MicrOmegas 3.3}~\cite{Belanger:2013oya} which we use to solve the relevant Boltzman equations numerically\footnote{
 It has recently been emphasized that 
the color structure of the final state for the $\tilde{f} \tilde{f}^\ast \rightarrow g g $ process forces a separate treatment for the odd and even angular momentum states \cite{ElHedri:2016onc}. This leads to
a different Sommerfeld factor for the $p$-wave compared to our implementation.
We have checked the numerical impact of this correction and find that it changes the predicted relic density by less than $1\%$ throughout the parameter space considered in this study. }.

  NLO calculations of the relic density have made significant progress in recent years and indicate that perturbative corrections can have a relevant impact on the annihilation rate \cite{Herrmann:2007ku,Baro:2007em,Harz:2012fz,Harz:2014tma,Harz:2014gaa,Harz:2016dql}. 
For stop coannihilation these corrections were found to change the relic density by $\approx 20 \%$~\cite{Harz:2014tma}. As expected the impact is even larger if Sommerfeld corrections are also included~\cite{Harz:2014gaa}. This is certainly larger than the observational uncertainty on the relic density and should be included in detailed studies of the MSSM. In the scenario at hand this correction can be absorbed via a shift of $\Delta m$ which does not have an appreciable effect of the qualitative features of the phenomenology. 

It has recently been emphasized that bound state formation may have an impact on dark matter freeze-out~\cite{Wise:2014jva,vonHarling:2014kha,Ellis:2015vaa,Kim:2016kxt,Liew:2016hqo,Mitridate:2017izz,Keung:2017kot}.
For squarks the non-Abelian structure of QCD leads to a partial accidental cancellation in the matrix element for bound state formation~\cite{Mitridate:2017izz} which somewhat diminishes its importance.
Once all electroweak annihilation channels are included in the freeze-out calculation, the impact of bound states was found to be at the $10\%$ level for all viable neutralino masses~\cite{Keung:2017kot}.  
We do not consider their effects here, but note that while the formation of bound states would allow for somewhat heavier dark matter consistent with the relic density constraint,  the qualitative picture would remain the same.

\begin{figure}[tb]
\begin{tabular}{cc}
\includegraphics[width=0.5\textwidth]{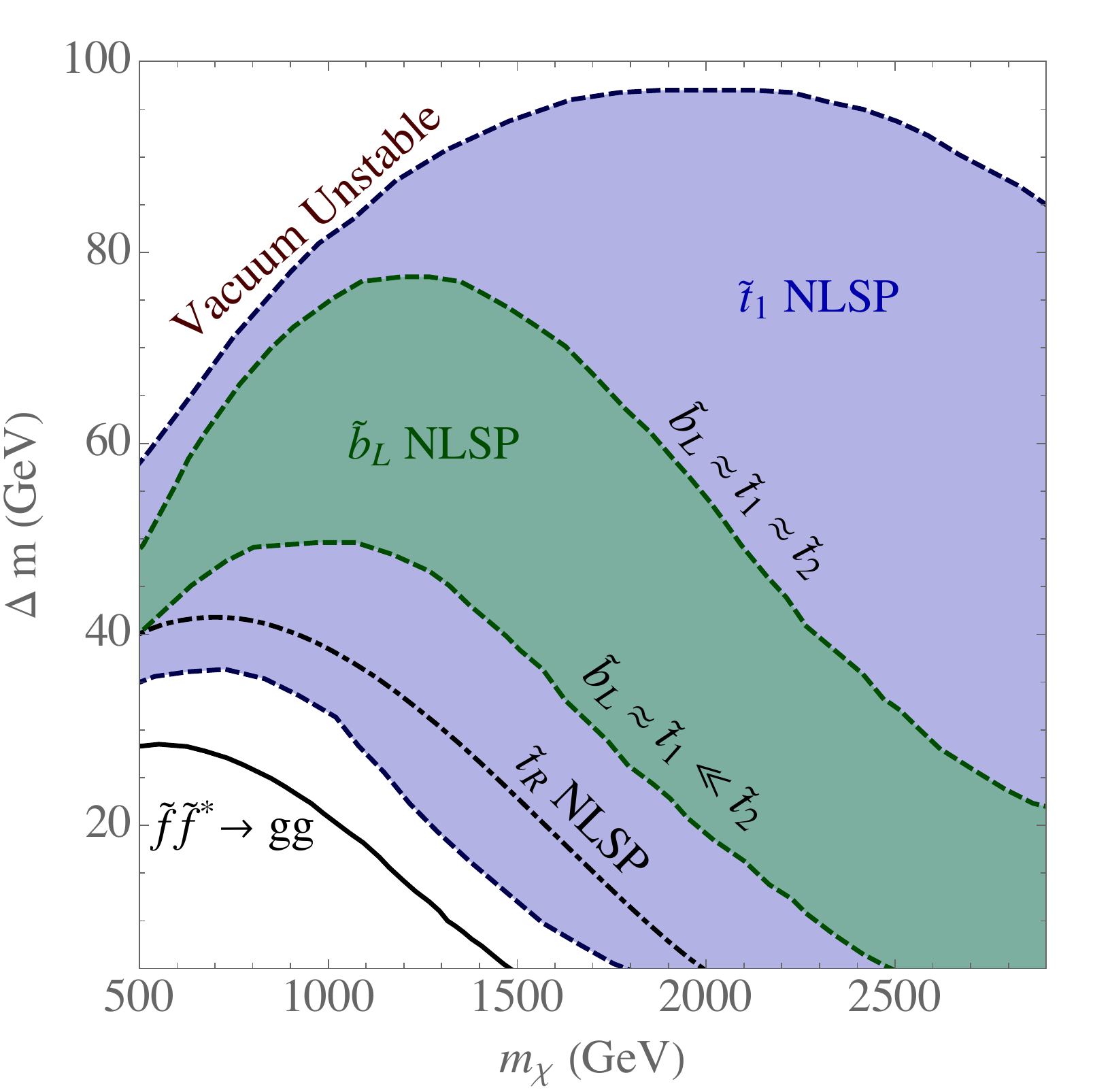}~~&~~
\includegraphics[width=0.5\textwidth]{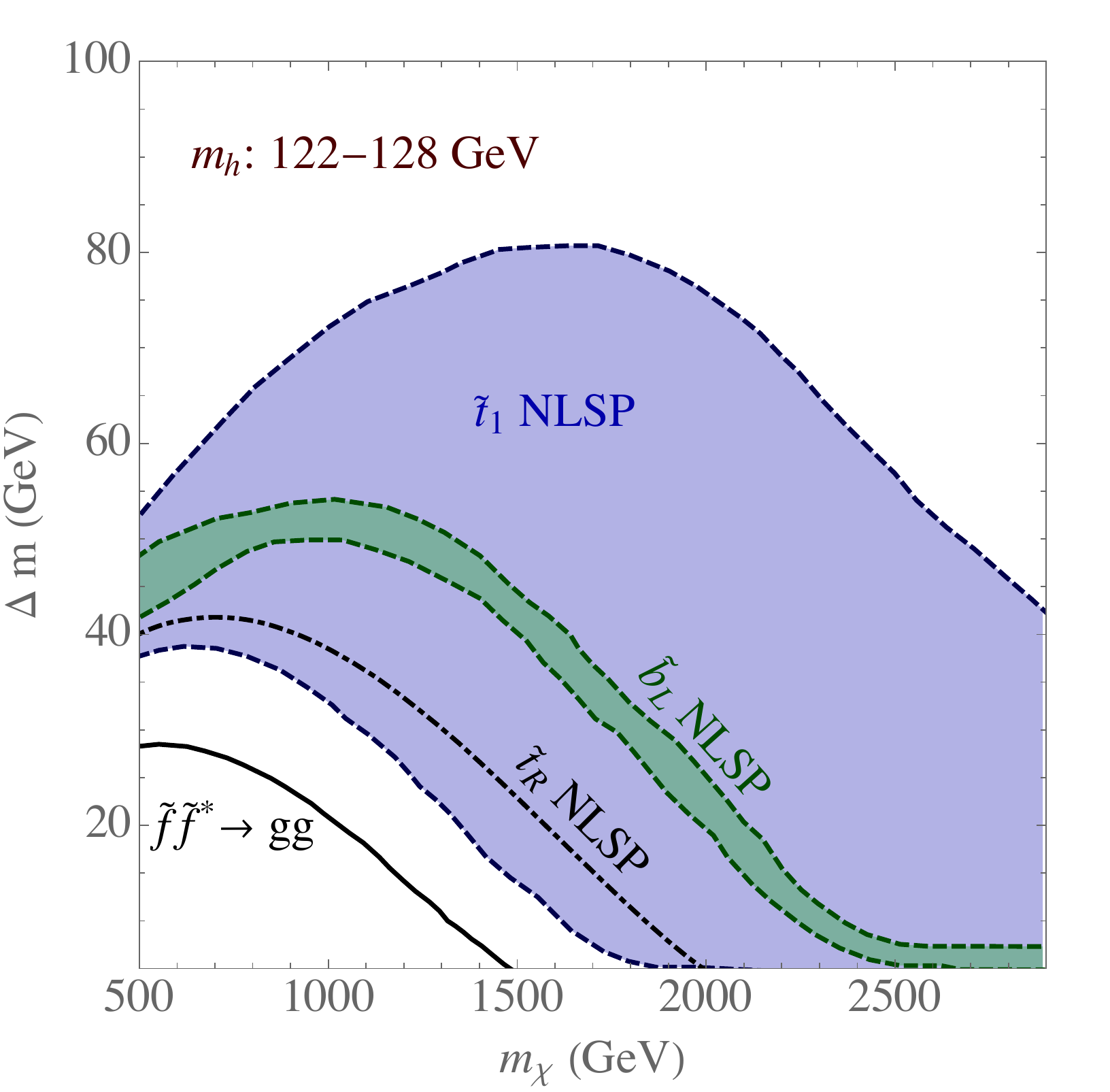}\\
(a) ~~&~~ (b)\\
\end{tabular}
\caption{ \label{f:deltam} {The mass difference $\Delta m$ required to obtain the correct relic density as a function of the LSP mass $m_{\chi}$. {\it Left:} All allowed masses, only constrained by the stability of the vacuum. {\it Right:} Requiring a Higgs mass between 122-128 GeV. The blue (green) band indicates the range of mass splittings for a $\tilde{t}_1$ ($\tilde{b}_L$) NLSP due to the different values of the stop mixing angle. The solid  black lines indicate the naive expectation if only QCD processes contribute to the relic density calculation. The dot-dashed black line marks the mass splitting consistent with only a right-handed light stop contributing to co-annhilations.}}
\end{figure}

In Fig.~\ref{f:deltam} we show the mass difference $\Delta m$ between $\chi$ and the NLSP where co-annihilations provide a viable mechanism for reproducing the dark matter density.  We have shown separately the regions where the left-handed sbottom is the NLSP and where the lightest stop is the NLSP.  To give guidance for the importance of various processes in the early universe we have shown two additional curves.  The first is the $\Delta m$ curve~(solid black line) if only the process $(\tilde{f} \tilde{f}^* \to gg)$ were active.  The 
true $\Delta m$ is always larger, showing the importance of other annihilation channels.  Another curve, labeled $\tilde{t}_R$~(dot-dashed black line), shows the mass splitting if the NLSP were a single pure right-handed stop.  In this case, there are no $X_{t}$ enhanced couplings to the Higgs boson, but the pure right-handed stop has the largest coupling to the bino~[cf. Eqs.~(\ref{ht1t1})-(\ref{bLx})].   In this case, the processes $\chi \chi \to t \bar{t}$, $\chi \tilde{t}_R \to t g/h$, $\tilde{t}_R \tilde{t}_R^* \to t\bar{t}/gg/hh$ are  relevant \cite{Ibarra:2015nca}.
 
The precise value of the mass splitting depends strongly on the mixing in the stop sector.  This mixing controls
both interactions with the Higgs boson and the bino, see~Eqs.~(\ref{ht1t1})-(\ref{t2xx}). The variation in these couplings due to different mixings explains the width of the band of consistent $\Delta m$ at each given value of the dark matter mass.   

In Fig.~\ref{f:deltam}, the allowed mass splitting for stop NLSPs broadens for higher dark matter masses.  Near $m_{\chi} \sim 500$ GeV, it populates a band from $\sim 35- 60 $ GeV, but in the multi-TeV range it can range from near degeneracy to mass splittings approaching 100 GeV.  The largest mass splittings require contributions from channels such as $(\tilde{t}_1 \tilde{t}_1^* \to hh)$ to be large -- realized by increasing  $X_t$~[cf. Eq.~(\ref{ht1t1})].
The cut-off in the maximal values of $\Delta m$ seen in the left panel 
results from the need for an $X_t$~$(A_t)$  so large as to make the vacuum unstable. The smallest $\Delta m$ obtained are somewhat smaller than the values consistent with a purely right handed stop NLSP.   
This is because perturbing away from the pure $\tilde{t}_R$ case, an important effect is that the $\tilde{t}_L$ admixture has a much smaller hypercharge and hence the coupling to $\chi$ is reduced, see Eq.~(\ref{t1xx}).  This suppresses diagrams that rely on this coupling, necessitating a smaller $\Delta m$ to maintain the thermal relic abundance.

In contrast, sbottom NLSPs never have $\Delta m$  exceeding  80 GeV, irrespective of the vacuum stability constraint. This is because for the sbottom to be the NLSP, $X_t$ cannot be too large (lest the $\tilde{t}_1$ mass be driven down by level repulsion).  In the sbottom NLSP case, the lightest stop will always be nearby, and for some parameter choices the heavier stop is also allowed to be degenerate. In fact the largest mass splitting obtained for the sbottom NLSP is precisely where both the stops are nearly degenerate with the sbottom, and dominant annihilation channels may include $\tilde{t}_{1,2}$ and $\tilde{b}_1$;  for example $(\tilde{f} \tilde{f}^* \to gg)$ or $(\tilde{t}_2/\tilde{t}_1 \tilde{b}_1 \to Wg)$. As the heavier stop decouples from the cosmology, the required mass splitting decreases and the smallest $\Delta m$ is obtained when only $\tilde{t}_1$ and $\tilde{b}_1$ contribute to co-annihilations.

The right panel of Fig.~\ref{f:deltam} shows the region that is carved out demanding consistency with the Higgs mass. Note that in particular the width of the sbottom NLSP region 
is significantly reduced. This can be understood by looking at the region in Fig.~\ref{fig:Higgs} where the sbottom is the NLSP: Only values of $s_t$ close to 0 and very large $m_{\tilde{t}_2}$ can accommodate both a sbottom NLSP and a consistent Higgs mass. This results in a well-defined cosmology for this scenario, and hence a well-defined $\Delta m$: values of the mass splitting are obtained close to the ones denoted by the bottom of the $\tilde{b}_1$ NLSP band in the left panel of Fig.~\ref{f:deltam}~($m_{\tilde{b}_L}\sim m_{\tilde{t}_1} \ll m_{\tilde{t}_2}$). Regarding the stop NLSP region,
some of the largest $\Delta m$ are eliminated 
as they require an $X_{t}$ so large that a large enough Higgs boson mass may not be achieved (see the right panel of Fig.~\ref{fig:Higgs}, particularly for $s_t > 0$).

While the masses of the stop NLSPs shown in Fig.~\ref{f:deltam} are in broad agreement with the latest LHC limits, it should be noted that the very lightest masses for a $\tilde{b}_1$ NLSP (approximately $m_{\chi} \lesssim 625$ GeV) are excluded by LHC searches, see Sec.~\ref{s.LHC}. The mass splittings shown in Fig.~\ref{f:deltam} represent important future targets for the LHC experiments.

\section{Direct Detection}\label{s.DD}
Having determined the regions of parameter space in which thermal freeze-out can account for the observed relic density, we now turn to  direct searches for dark matter.
\begin{figure}[tb]
\begin{tabular}{ccc}
 \includegraphics[width=0.25\textwidth]{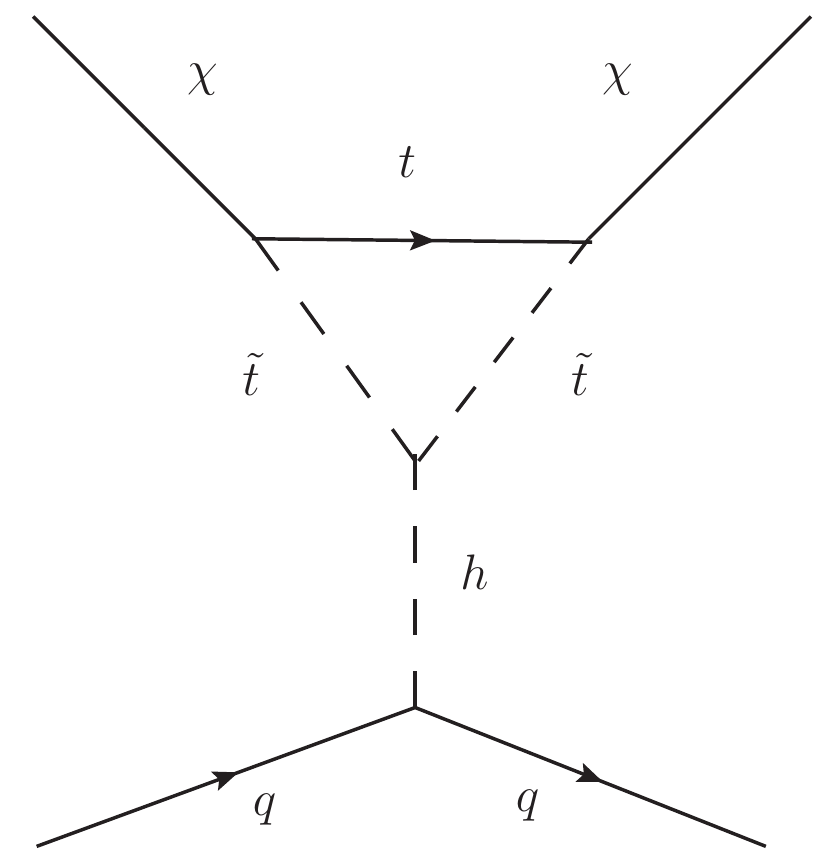}~~ & ~~\includegraphics[width=0.25\textwidth]{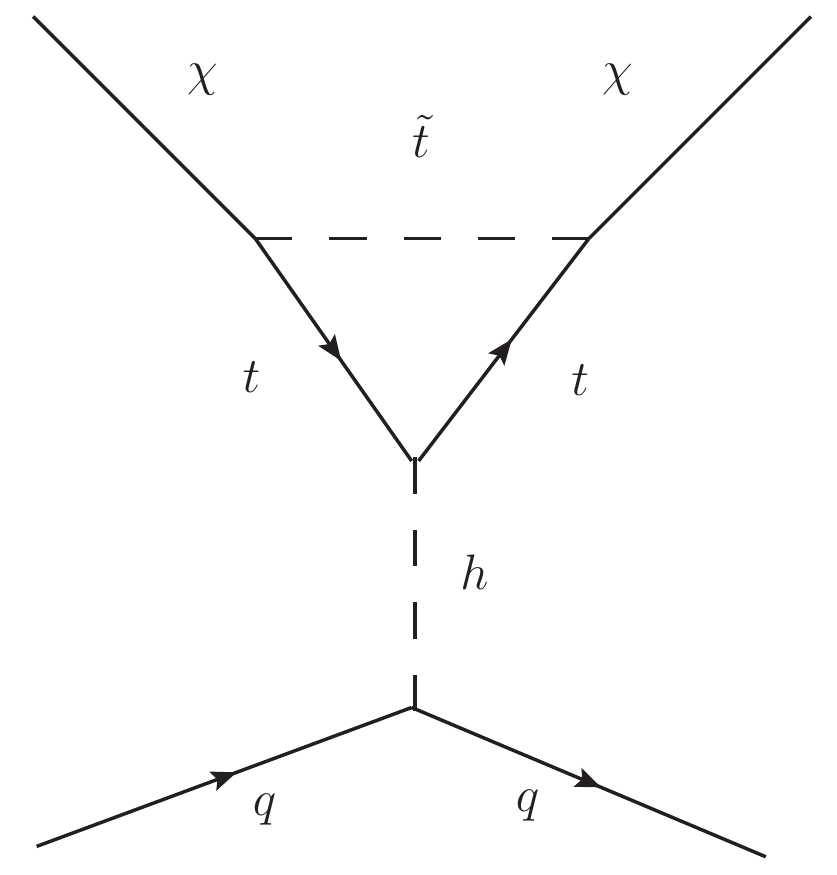}~~ & ~~
 \includegraphics[width=0.27\textwidth]{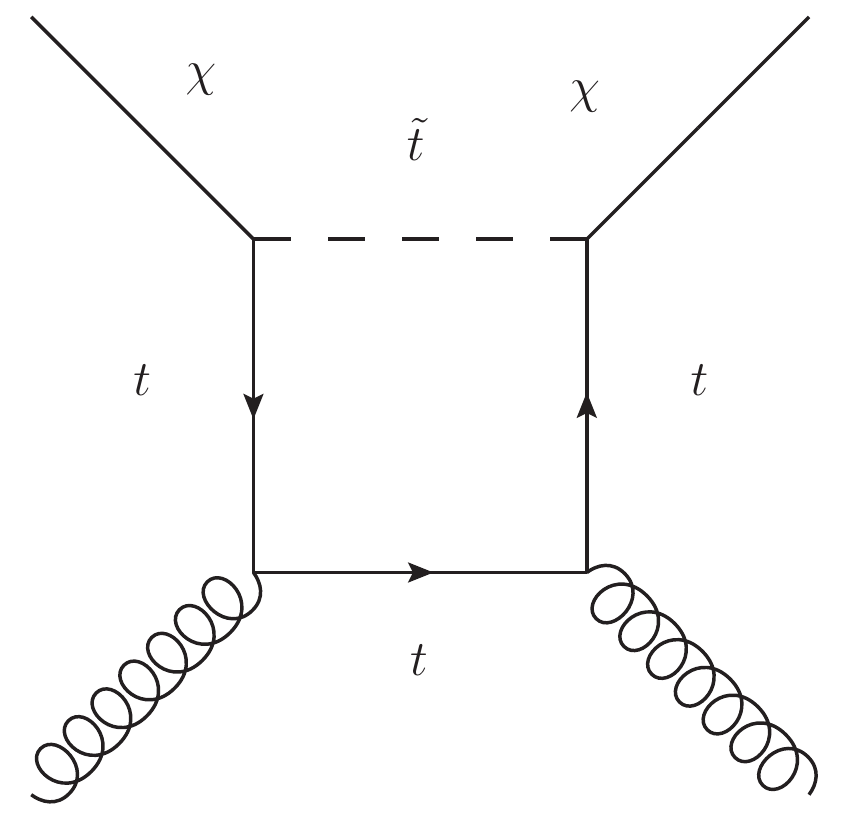} \\
 (a) & (b) & ~~~(c)
\end{tabular}
\caption{ \label{fig:DDdiagrams} {   Triangle $\chi\chi h$~[(a) and (b)] and  box $gg\chi\chi$~(c) diagrams contributing to the dark matter nucleon coupling.}}
\end{figure}
The bino dark matter candidate considered here has a vanishing direct detection cross section at tree level.
A process that can generate a direct detection cross section arises from the loop-induced coupling of the dark matter with the Higgs boson. Such an effective $\chi\chi h$ coupling is generated by triangle diagrams with tops and stops in the  loop, see Figs.~\ref{fig:DDdiagrams}a and \ref{fig:DDdiagrams}b. This effective coupling has been calculated in Ref.~\cite{Djouadi:2001kba} and  re-derived by us using the low energy Higgs theorem~\cite{Kniehl:1995tn}.\footnote{Loops with the $\tilde{t}$ will also contribute to the Higgs coupling to gluons. This effect is always subleading and  we do not include it in the following qualitative discussion. Our numerical calculations take the effect into account. For a discussion of the full SUSY-QCD corrections to direct detection see for instance \cite{Klasen:2016qyz}}   In addition to the Higgs triangle-diagrams there are also box-diagrams with $\tilde{t}_{1,2}/t$ and $\tilde{b}_1/b$ in the loop which induce an effective 
coupling between the bino and the gluon content in the nucleus~(see Fig.~\ref{fig:DDdiagrams}c as an example diagram). The full loop result for this contribution is available in the literature~\cite{Drees:1993bu} and is already included in the MSSM implementation of {\tt MicrOmegas}.

\begin{figure}[t]
\includegraphics[width=0.7\textwidth]{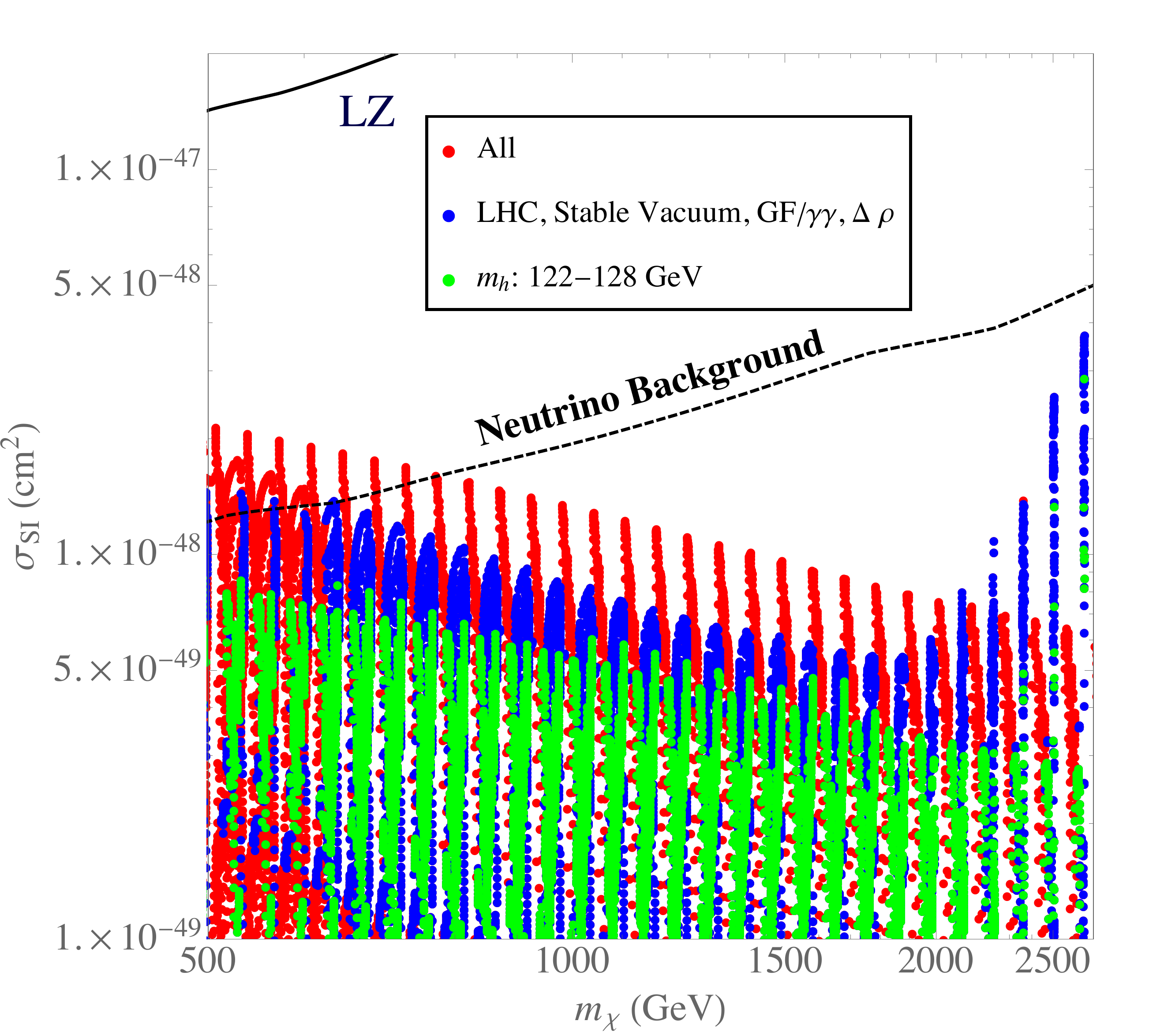}
\caption{ \label{f:DD} {The spin-independent direct detection cross section $\sigma_{SI}$ as a function of $m_\chi$. In red are all points in our scan, which include points that violate the vacuum stability constraint on $X_t$ by up to 50\%.  Blue points satisfy indirect constraints ($\delta \rho$, Higgs signal strength, and vacuum stability), see Sec.~\ref{s.Higgs}, as well as direct searches at the LHC, see Sec.~\ref{s.LHC}.  Green points also realize a Higgs mass in the window 122 GeV $< m_{h} <$ 128 GeV, see Sec.~\ref{s.Mass}.  The expected sensitivity of the future LZ experiment and the ultimate sensitivity achievable in light of the irreducible neutrino background are indicated by a solid and dashed line respectively, taken from Ref.~\cite{Feng:2014uja}. }}
\end{figure}

We show the direct detection cross section as a function of the neutralino mass  in Fig.~\ref{f:DD} considering different constraints.  First, in red, we show all points in our scan, which include $X_{t}$ that would violate the vacuum stability constraint by as much as $50\%$.  Next, in blue, we show points consistent with the observed Higgs signal strength (gluon fusion rate), EWPO ($\delta \rho$), and a metastable vacuum with sufficiently long lifetime.  Because all points shown have $m_{\chi} > 500$ GeV, they are consistent with current stop searches \cite{CMS-PAS-SUS-16-049}.  In addition, we ensure that the points are consistent with the recent sbottom searches \cite{CMS-PAS-SUS-16-032}, which eliminates some points with mass between 500 and 625 GeV where the sbottom was nearly degenerate with the LSP.  Finally, in green, we show points that also yield a Higgs mass 122 GeV $< m_{h} <$ 128 GeV.  Once the Higgs  mass is imposed, all points lie well below future experimental sensitivity, and indeed below the neutrino floor.  We can understand the behavior of the direct detection in some detail by examining the relevant amplitudes.
 
At the nucleon level the direct detection cross section is given by:
\be
\sigma_{SI}^n = \frac{4m_\chi^2m_n^4\, \mathcal{A}^2}{\pi (m_\chi+m_n)^2}\, ,
\ee
where $n$ specifies the nucleon (i.e. the neutron or the proton) and $\mathcal{A}$ is the amplitude which can be separated into a  Higgs exchange contribution and the gluon box contribution
\be
\mathcal{A}=\mathcal{A}^h+\mathcal{A}^g.
\ee
We discuss each of these contributions in turn, attempting to understand how large an amplitude can be realized for each.  

The gluon box diagram is suppressed by the relevant fermion mass and hence can be enhanced when the sbottom is the NLSP and the mass splitting is comparable to $m_b$. In this case, $\mathcal{A}^g$ is  well approximated by 
\begin{eqnarray}\label{e:glubox}
\mathcal{A}^g &\simeq& \frac{g^2 t_w^2}{36}\frac{1}{m_\chi \left(m_b+\Delta m\right)^2}\left[ \frac{f_{T_g}^{n}}{270}+\frac{13}{240 \pi} G_\alpha^b\right] \nonumber \\              
                        &\approx &1.5 \times 10^{-11} \, \mathrm{GeV}^{-3} \left(\frac{2 \, \mathrm{TeV}}{m_\chi}\right)\left( \frac{m_b+20~ \mathrm{ GeV} }{m_b+\Delta m}\right)^2,
\end{eqnarray}
where we use default values from {\tt MicrOmegas} for $f_{Tg^b}^n\simeq 0.98$ and $G_\alpha^b\simeq 0.1$. 

The Higgs amplitude receives contributions from diagrams where the Higgs boson couples to the $t$ as well as to both the stops $\tilde{t}_{1,2}$.   It can be decomposed as 
\begin{eqnarray}
\label{eqn:HiggsExchangeDD}
\mathcal{A}^h&=&\frac{g^2t_w^2}{24 \pi^2}\frac{m_t}{v^2 m_h^2}~ \mathcal{A}^h_{t\tilde{t}}~\sum_q f_q^n  
,  \qquad\quad  f_q^n = \left \{\begin{array}{c c}  f_{Tq}^n,& \quad \quad  q=\{u,d,s\} \\
  \frac{2}{27} f_{Tg}^n, & \,\;\;\quad q=\{c,b,t\} 
    \end{array} \right. \, \nonumber\\
   & \approx&  2.7 \times 10^{-11} \mathrm{GeV}^{-3} ~ \mathcal{A}^h_{t\tilde{t}},
       \end{eqnarray}
where $ \mathcal{A}^h_{t\tilde{t}} = \left( \mathcal{A}^h_t+\mathcal{A}^h_{\tilde{t}_{1,1}}+\mathcal{A}^h_{\tilde{t}_{1,2}}+\mathcal{A}^h_{\tilde{t}_{2,2}} \right) $, $f_{Tg}^n = (1-\sum_{u,d,s}f_{Tq}^n)$ parametrizes the gluon contribution to the nucleon mass,  $f_{Tq}^n$ are the contributions from the light quark species $u,d$ and $s$, and  $v =246$ GeV is the electroweak vacuum expectation value. We use the default values in {\tt MicrOmegas} which  lead to $\sum_q f_q^n \simeq 0.28$.

As we showed in the previous section,  the mass splitting necessary to reproduce the thermal relic density is much smaller than the top mass independently of whether the NLSP is the stop or the sbottom. In this case, taking advantage of the expressions for the couplings Eqs.~(\ref{ht1t1})--(\ref{bLx}), and using the relevant limit for the loop integrals, the various contributions to the amplitudes are well approximated by
\bea
 \mathcal{A}_t^h&\simeq& \frac{s_{2t}}{2}\left(1-\frac{m_{\tilde{t}_1}^2}{m_{\tilde{t}_2}^2}\right)+\left(s_t^2+\frac{c_t^2}{16}\right) \frac{m_t}{m_{\tilde{t}_1}}+\left(c_t^2 +\frac{s_t^2}{16}\right)\frac{m_t m_{\tilde{t}_1}}{m_{\tilde{t}_2}^2} \;,\label{eq:t}\\
 \mathcal{A}^h_{\tilde{t}_{1,1}} &\simeq & \frac{m_t}{m_{\tilde{t}_1}}\left( 1-\frac{X_t^2}{m_{\tilde{t}_2}^2-m_{\tilde{t}_1}^2} \right)\left[ 2 \left(s_t^2+\frac{c_t^2}{16}\right) -s_{2t} \frac{m_t}{m_{\tilde{t}_1}}\right] \;, \label{eq:t1}\\
 \mathcal{A}^h_{\tilde{t}_{1,2}} &\simeq & -\frac{X_t}{m_{\tilde{t}_2}} c_{2t}\left( -\frac{15}{16}s_{2t}+c_{2t} \frac{m_t}{m_{\tilde{t}_1}} \right) \;, \label{eq:t1t2}\\
  \mathcal{A}^h_{\tilde{t}_{2,2}} &\simeq & \frac{m_t m_{\tilde{t}_1}}{m_{\tilde{t}_2}^2}\left( 1+\frac{X_t^2}{m_{\tilde{t}_2}^2-m_{\tilde{t}_1}^2} \right)\left[2 \left(c_t^2 +\frac{s_t^2}{16}\right)+ s_{2t}  \frac{m_t}{m_{\tilde{t}_1}}\right]\;.\label{eq:t2}
\eea
Unless  $X_t \gg m_{\tilde{t}_{1,2}}$, the above can be seen to be at most $\mathcal{O}(1)$.

Taking into account these
potentially important contributions, we can write down how the direct detection cross section scales ($m_\chi \sim m_{\tilde{t}_1}\sim m_{\tilde{b}_1}$)  
\be\label{e:SIapprox}
\sigma_{SI} \sim 4 \times 10^{-48} \mathrm{cm^2}\left[ 0.15\left( \frac{2~ \mathrm{TeV}}{m_\chi}\right)\left(\frac{m_b +20~ \mathrm{GeV}}{m_b+\Delta m}\right)^2 + \mathcal{A}^h_{t\tilde{t}}  \right]^2,
\ee
where $\Delta m$ is the splitting with the sbottom. It should be noted that this expression will breakdown at smaller cross sections, in part, due to the omission of stop loop contributions to the gluon amplitude.

We now discuss how $ \mathcal{A}^h_{t\tilde{t}}$ behaves in various limits to gain further insight into the expected size of the direct detection cross section.   Naively, it appears that the  cross section may be driven arbitrarily large for maximally mixed stops, $s_{2t} \simeq1$  by taking  $X_t$ large~(hence $m_{\tilde{t}_2} \gg m_{\tilde{t}_1}$), where the $\tilde{t}_1$ contribution completely dominates $\mathcal{A}^h_{t\tilde{t}}$ giving $\mathcal{A}^h_{t\tilde{t}} \sim X_t/ 2m_{\tilde{t}_1}$. However,  constraints both from vacuum stability and the Higgs mass prevent going too far into this regime.

Once we impose the Higgs mass constraint it is possible to make detailed statements about the Higgs mediated amplitude.  
As can be seen from Fig.~\ref{fig:Higgs}, for any given mixing angle, generally the largest values of $X_t$ or $m_{\tilde{t}_2}$ consistent with the Higgs boson mass are close to the largest values allowed by the stability of the vacuum.  It is useful to consider different mixing angles separately.  First, we examine the case of significantly mixed stops.   Here, the proper Higgs boson mass is obtained for values of $m_{\tilde{t}_2}$ not too much heavier than $m_{\tilde{t}_1}$ and approximately $|X_t| \sim 2~ m_{\tilde{t}_2}$. The contribution to the amplitude from the stops can then be approximated as  $\mathcal{A}^h_{t\tilde{t}} \sim  m_t/ m_{\tilde{t}_1}\left[ 3-X_t/m_{\tilde{t}_2}\left(2+X_t/m_{\tilde{t}_2}  \right)  \right]\sim 0.26$ for positive $s_t$~($X_t<0$) and $m_{\tilde{t}_1}$ = 2 TeV. For negative $s_t$ ($X_t >0$), this contribution can be approximately a factor of 2 larger in magnitude than the $s_t >0$ case, but with a negative sign.  This negative sign yields destructive interference with the subdominant gluon contribution.   We now turn to the case of small, but finite mixing, ($s_t\sim1$ or 0).  In this case, a consistent Higgs mass is  obtained when $m_{\tilde{t}_1} \ll m_{\tilde{t}_2}$ and $|X_t| \sim m_{\tilde{t}_2}$. For a predominantly right-handed light stop,  we find $\mathcal{A}^h_{t\tilde{t}} \sim m_t/m_{\tilde{t}_1} \left[ 3-X_t/m_{\tilde{t}_2}\left(1+2X_t/m_{\tilde{t}_2}  \right) +X_t m_{\tilde{t}_1}/m_{\tilde{t}_2}^2\left(-1+2X_t/m_{\tilde{t}_2}  \right)  \right]\sim 0.35$ for $m_{\tilde{t}_1}=2$ TeV,  $m_{\tilde{t}_2}=3$  TeV  and positive $s_t~(X_t<0)$. As can be seen from Eqs.~(\ref{eq:t})-(\ref{eq:t2}), most of the contributions to the direct detection cross section for a purely left-handed stop, close to $c_t\sim1$, are suppressed by factors of 1/16 compared to the right-handed approximation owing to the smaller hypercharge of the left-handed multiplet. However, there is an unsuppressed contribution from $\mathcal{A}^h_{\tilde{t}_{1,2}}$, and as long as the mixing angle is not too small, it tends to dominate. It leads to an approximate amplitude $\mathcal{A}^h_{t\tilde{t}} \sim - m_t X_t/m_{\tilde{t}_1} m_{\tilde{t}_2}\sim 0.09$ for negative $X_t$ and $m_{\tilde{t}_1}=$ 2 TeV. Given these considerations, and demanding consistency with the Higgs mass, it is clear that unless the mass splitting between $\chi$ and $\tilde{b}_L$ is
$\sim m_b$, which happens for $m_\chi\gtrsim 2.5$ TeV, the direct detection cross section for a predominantly left-handed stop is much smaller than a right-handed stop.
Due to the destructive interference of the stop contributions with the gluon contribution for $s_t<0$, the maximal cross section, in the absence of very small mass-splitting with the sbottom, consistent with the Higgs mass is obtained for a dominantly right-handed light stop with $s_t>0$. This maximal value can be roughly approximated by  $\sigma_{SI} \sim 10^{-49}\left( 1+ 3 \tilde{t}_1/2\tilde{t}_2\right)^2\left(2 ~\mathrm{TeV}/ \tilde{t}_1\right)^2\mathrm{cm}^2$, a value well below the neutrino floor.

 Finally, a comment regarding the direct detection prospects for a mixed $\tilde{b}$ co-annihilation scenario is in order.  We have set $X_{b}$ identically zero throughout, in part for simplicity, and in part because a mixing in the sbottom sector is not motivated by the Higgs boson mass.  Relaxing this assumption would allow the lightest sbottom to be an admixture of the right and left-handed states.  In cases where the sbottom is the co-NLSP, this enhances the direct detection cross section with respect to Eq.~(\ref{e:SIapprox}) due to the larger hypercharge of the $\tilde{b}_R$. To get a feel for the maximum size of this effect, we consider the limiting case of a pure $\tilde{b}_R$ NLSP.
 The $\tilde{b}_L$ and the entire stop sector can be substantially heavier than the NLSP.  In this case the thermal expectation for $\Delta m $ found in Ref.~\cite{Keung:2017kot} is rather similar to the expectation for $(\tilde{f} \tilde{f}^* \rightarrow g g)$ shown in Fig.~\ref{f:deltam}. Note that this is substantially smaller than the expectation for a purely left-handed sbottom NLSP and hence will lead to significant enhancement of the gluon box contribution, Eq.~(\ref{e:glubox}).  The corresponding direct detection cross section has recently been studied in Ref.~\cite{Berlin:2015njh}, and for a thermal mass splitting, as read from our Fig.~\ref{f:deltam}, their results indicate an expected scattering rate above the neutrino floor throughout the allowed parameter space.  However,
LZ is only expected to be sensitive to the very smallest mass splittings, corresponding to 1300 GeV $\lsim  m_{\chi} \lsim$ 1700 GeV.

\section{Conclusion}\label{s.conc}

We have revisited the stop co-annihilation scenario
wherein dark matter freeze-out, stop-induced corrections to the Higgs boson mass as well as an irreducible direct detection cross section mediated by stop/sbottom loops are interrelated. After accounting for collider limits on light stops and sbottoms we find that the constraints on the stop sector from the stability of the vaccuum  are more stringent than EWPOs and the Higgs signal strength measurements. The region of stop parameters consistent with the Higgs mass is generally unconstrained by any of the above listed indirect constraints, however there may be slight tension between the largest values of $X_t$ consistent with the Higgs mass and the stability of the vacuum. 
 
 We analyze the irreducible, stop/sbottom-loop induced direct detection cross section in detail taking into account the mass splittings motivated by cosmology. Unfortunately, we find that this minimal direct detection rate generally falls well below the neutrino floor, particularly once the Higgs mass is taken into account.
It should be kept in mind, however, that even a small amount of bino-Higgsino mixing can induce substantial spin-independent  neutralino-nucleus scattering at tree-level. In this case the rate could be large, even going up to the current limit, but the direct connection between the cosmological origin of dark matter in the early Universe and direct detection rates would be lost. 

In light of the potentially challenging direct detection  situation, collider searches may be 
the best available probes for this scenario.
The mass splitting between the  $\tilde{t}$~(or~$\tilde{b})$  and the neutralino is small in the co-annihilation region, and naively,  absent a detailed model of supersymmetry breaking it is not clear why this should be so.  However, given that it produces the proper thermal dark matter relic density consistent with increasingly strong direct detection bounds, this mass spectrum is extremely well motivated.
Therefore the region in the $\Delta m$ vs. $m_{\chi}$ plane shown in Fig.~\ref{f:deltam} represents an important target.  Dedicated LHC searches for such compressed spectra will be crucial in testing the co-annihilation scenario.

\section*{Acknowledgements}
AP thanks D. Morrissey and N. Blinov for communicatons regarding charge color breaking vacuua. We would like to thank K. Zurek, A. Berlin and S. Wild for discussions related to direct detection and the authors of~\cite{ElHedri:2016onc} for valuable comments regarding Sommerfeld corrections. NRS thanks C. Wagner for discussions regarding Higgs phenomenology.
This material is based upon work supported by the U.S. Department of Energy, Office of Science, Office of High Energy Physics under Award Number DE-SC0007859. 
AP and NS thank the Aspen Center for Physics and the NSF Grant \#1066293 for hospitality.

\bibliography{Stops_AP}
\bibliographystyle{utphys}
\end{document}